\documentclass[amsmath,prl,aps,onecolumn,superscriptaddress,reprint]{revtex4-1}
\usepackage[english]{babel}
\PassOptionsToPackage{english}{babel}
\usepackage{lineno}
\usepackage[utf8]{inputenc}
\usepackage{color}
\usepackage{amssymb,graphicx,xcolor,units,hyperref}
\usepackage{booktabs}
\usepackage{soul}
\usepackage{float}

\usepackage{hyperref}
\usepackage{makecell}
\hypersetup{
    bookmarks=false,         % show bookmarks bar?
    unicode=false,          % non-Latin characters in AcrobatÕs bookmarks
    pdftoolbar=false,        % show AcrobatÕs toolbar?
    pdfmenubar=true,        % show AcrobatÕs menu?
    pdffitwindow=false,     % window fit to page when opened
    pdfstartview={FitH},    % fits the width of the page to the window
    pdftitle={Detecting induced $p \pm ip$ pairing at the Al-InAs interface with a quantum microwave circuit},    % title
    pdfauthor={D. Phan et al},     % author
    pdfsubject={},   % subject of the document
    pdfcreator={},   % creator of the document
    pdfproducer={}, % producer of the document
    pdfkeywords={}, % list of keywords
    pdfnewwindow=true,      % links in new window
    colorlinks=true,       % false: boxed links; true: colored links
    linkcolor=black,          % color of internal links (change box color with linkbordercolor)
    citecolor=blue,        % color of links to bibliography
    filecolor=magenta,      % color of file links
    urlcolor=blue           % color of external links
}

\begin{abstract}
\end{abstract}

\begin{document}
\selectlanguage{english}
\title{Detecting induced $p \pm ip$ pairing at the Al-InAs interface with a quantum microwave circuit}

\author{D. Phan}
\affiliation{IST Austria, Am Campus 1, 3400 Klosterneuburg, Austria}
\author{J. Senior}
\affiliation{IST Austria, Am Campus 1, 3400 Klosterneuburg, Austria}
\author{A. Ghazaryan}
\affiliation{IST Austria, Am Campus 1, 3400 Klosterneuburg, Austria}
\author{M. Hatefipour}
\affiliation{Department of Physics, New York University, New York, NY, 10003, USA}
\author{W.~M.~Strickland}
\affiliation{Department of Physics, New York University, New York, NY, 10003, USA}
\author{J. Shabani}
\affiliation{Department of Physics, New York University, New York, NY, 10003, USA}
\author{M. Serbyn}
\affiliation{IST Austria, Am Campus 1, 3400 Klosterneuburg, Austria}
\author{A. P. Higginbotham}
\affiliation{IST Austria, Am Campus 1, 3400 Klosterneuburg, Austria}
\email{andrew.higginbotham@ist.ac.at}

\date{\today}

\begin{abstract}
Superconductor-semiconductor hybrid devices are at the heart of several proposed approaches to quantum information processing, but their basic properties remain to be understood.
We embed a two-dimensional Al-InAs hybrid system in a resonant microwave circuit, probing the breakdown of superconductivity due to an applied magnetic field.
We find a fingerprint from the two-component nature of the hybrid system, and quantitatively compare with a theory that includes the contribution of intraband $p \pm i p$ pairing in the InAs, as well as the emergence of Bogoliubov-Fermi surfaces due to magnetic field.
Separately resolving the Al and InAs contributions allows us to determine the carrier density and mobility in the InAs.
\end{abstract}

\maketitle

Hybrids of superconducting and semiconducting materials are under investigation as platforms for integrated superconducting devices~\cite{rusen_gannbn_2018,wen_epitaxial_2021}, superconducting qubits~\cite{larsen_semiconductor-nanowire-based_2015,delange_realization_2015,casparis_gatemon_2016,casparis_superconducting_2018}, and engineered $p$-wave superconductivity~\cite{fu_superconducting_2008,fujimoto_topological_2008,zhang_px_2008,flensberg_engineered_2021}.
Hindering progress towards these goals, basic semiconductor properties such as carrier density, mobility, and induced pairing are currently inaccessible because the superconductor acts as a perfectly conductive shunt.
This problem is especially acute in the ongoing effort to conclusively identify Majorana modes~\cite{mourik_signatures_2012,lee_spin-resolved_2014,lutchyn_majorana_2018,chen_ubitquitous_2019,yu_non-majorana_2021,valentini_non-topological_2021,hart_controlled_2017,ren_topological_2019,fornieri_evidence_2019,dartiailh_phase_2021}.
Due to the bulk-boundary correspondence, the presence of these modes should be controlled by bulk, as yet undetermined, semiconductor parameters.
In particular, depending on parameter values, application of a magnetic field can result in transitions to the normal state~\cite{chandrasekhar_note_1962,clogston_upper_1962,tewari_topologically_2011}, partial Bogoliubov-Fermi surfaces~\cite{yuan_zeeman-induced_2018}, gapless $p_x$ phases~\cite{alicea_majorana_2010}, or chiral $p$-wave phases with Majorana modes~\cite{sato_topological_2009,lee_proposal_2009,sau_non-abelian_2010,alicea_majorana_2010,lutchyn_majorana_2010,oreg_helical_2010}.

In this Letter, we experimentally study induced superconductivity in a two-dimensional Al-InAs hybrid system using a resonant microwave circuit.
Above a characteristic field we discover anisotropic suppression of superfluid density and enhanced dissipation, consistent with a picture of two fully gapped, intraband $p \pm i p$ superconductors transitioning to partial Bogoliubov-Fermi surfaces. 
Observation of this transition allows for the characterization of key system properties such as induced pairing, carrier density, and carrier mobility.	
We therefore demonstrate the first evidence of two-dimensional induced $p$-wave pairing, the emergence of Bogoliubov-Fermi surfaces, and a general method for characterizing otherwise invisible properties of superconductor-semiconductor hybrid devices.

\begin{figure}[b]
	\centering
	  \includegraphics[scale=1]{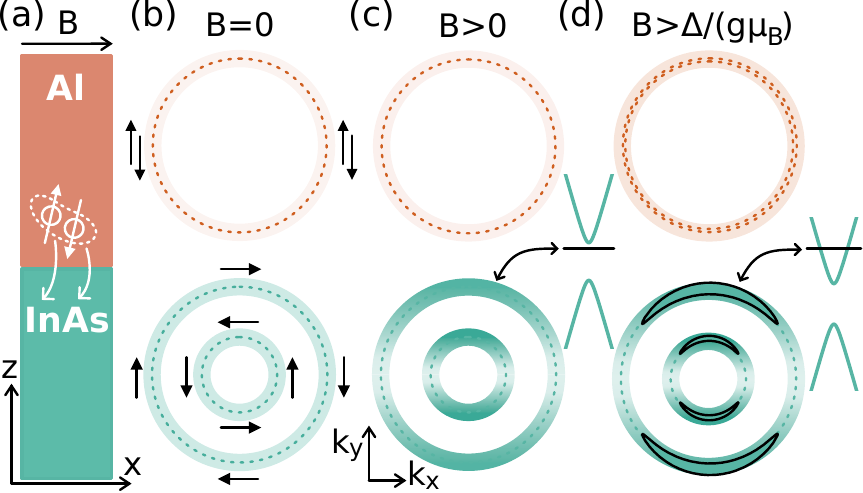}
		\caption{
		(a) Physical picture of proximity effect between Al in InAs.
		Field direction, $B$, indicated.
		(b) Al has a spin-degenerate Fermi surface gapped by superconductivity (orange). InAs has two spin-orbit coupled Fermi surfaces with intraband $p \pm i p$ pairing (green). %Fermi surfaces not drawn to scale.
		(c) Magnetic field anisotropically reduces InAs gap (color intensity)
		Hyperbolas indicate quasiparticles dispersion. Black line indicates chemical potential.
		(d) For $B>\Delta/(g \mu_B)$ the InAs gap closes in isolated regions, forming connected arcs of zero-energy electron-like and hole-like quasiparticles, known as Bogoliubov-Fermi surfaces.
		}
	  \label{fig:0}
  \end{figure}

\begin{figure*}
	\centering
	  \includegraphics[scale=1.0]{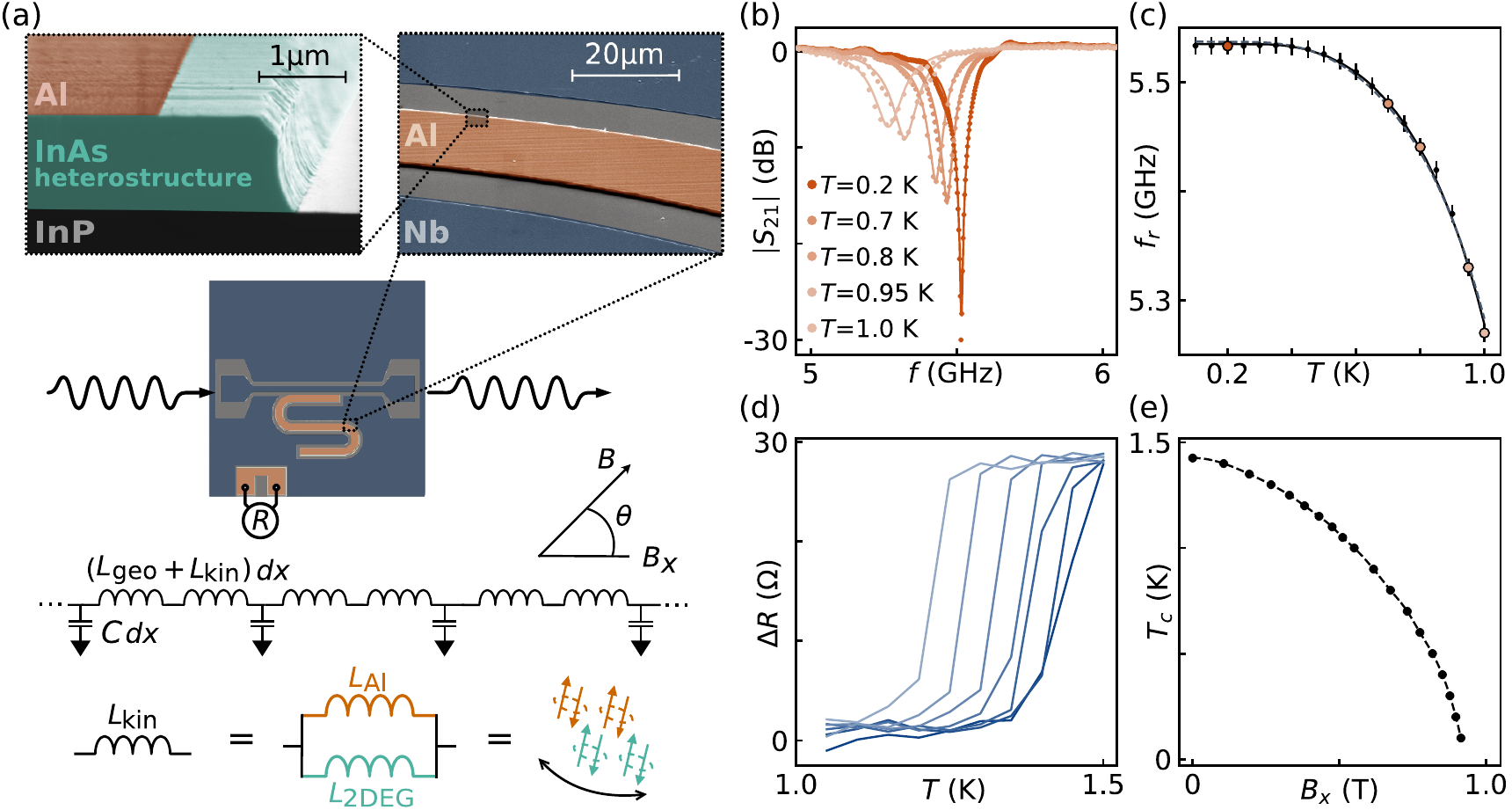}
		\caption{
		(a) False-color scanning-electron micrographs of an example Al-InAs device.
		The InAs heterostructure houses a two-dimensional electron gas (top).
		Schematic of the chip layout with microwave resonator, transport device allowing for measurement of resistance $R$, magnetic-field angle $\theta$ (middle).
		Transmission-line model with geometric inductance $L_\mathrm{geo}$ and kinetic inductance $L_\mathrm{kin}$, which receives a contribution from the Al and from the InAs.
	  (b) Microwave transmission $S_{21}$ as a function of frequency, measured for different cryostat temperatures.
	  Solid lines are fits using method of Ref.~\cite{probst_efficient_2015}.
	  (c) Resonant frequency $f_r$ extracted from (a) versus cryostat temperature $T$, colored points match like-colored traces in (b). 
	  Solid line is a fit to $s$-wave theory including disorder ($c_p$=0), dashed line is a fit to the two-component model.
	  (d) Resistance $R$ vs temperature $T$. Curves from right to left have $B_x$ uniformly increasing from 0 to $0.36~\mathrm{T}$.
	  A $15~\mathrm{\Omega}$ overall offset has been subtracted from the data.
	  (e) Critical temperature as a function of $x$-oriented magnetic field $B_x$.
	  Points joined by an interpolating function, used for smoothly estimating $T_c(B)$ in the pair-breaking numerical fit. }
	  \label{fig:1}
  \end{figure*}

The basic picture of proximity effect in Al-InAs is presented in Fig.~\ref{fig:0}.
An aluminum layer with a spin-degenerate Fermi-surface is strongly coupled to a high-mobility InAs two-dimensional electron gas.
InAs has a pair of spin-orbit coupled Fermi surfaces which results in $p \pm i p$ intraband pairing of the form $\Delta (k_x \pm i k_y)/|k|$ for pure Rashba spin-orbit interaction, where $k$ is the momentum at the Fermi surface labeled by $\pm$~\cite{fu_superconducting_2008,alicea_majorana_2010}.
This pairing holds a special importance because a state with single, chiral $p_x+i p_y$ pairing is topologically nontrivial, and therefore capable of hosting Majorana modes~\cite{read_paired_2000}.
Application of an in-plane magnetic field probes the nature of the induced pairing.
For weak spin-orbit coupling, interband $s$-wave pairing quickly emerges~\cite{alicea_majorana_2010,potter_engineering_2011}, and the system eventually transitions to an isotropic normal state~\cite{tewari_topologically_2011}.
In contrast, strong spin-orbit coupling makes the $p \pm i p$ pairing robust. 
Magnetic field then generates anisotropic suppression of the induced gap, eventually causing the emergence of Bogoliubov-Fermi surfaces~\cite{yuan_zeeman-induced_2018}.
Bogoliubov-Fermi surfaces, however, may be subject to an instability that was explored in related systems~\cite{liu_interior_2003,wu_superfluidity_2003,forbes_stability_2005,agterberg_bogoliubov_2017,setty_bogoliubov_2020}.
Thus the presence of $p\pm ip$ pairing qualitatively affects the response of superconductivity to in-plane magnetic fields, motivating the present study.

In order to probe the effect of magnetic field on induced superconductivity, we construct a half-wave coplanar waveguide resonator with a center pin made from an Al-InAs superconductor-semiconductor heterostructure~\cite{shabani_two-dimensional_2016,kjaergaard_coupling_2016,william_superconducting_2019}, shown in Fig.~\ref{fig:1}(a) with more material details in~\cite{supplement}.
The resonant frequency of this circuit is altered by the condensate kinetic inductance, which is inversely proportional to superfluid density $ \rho_\mathrm{SF}$~\cite{annunziata_tunable_2010,driessen_strongly_2012}.
The emergence of Bogoliubov-Fermi arcs is expected to deplete the contribution of InAs to $\rho_\mathrm{SF}$, and thus alter the circuit resonant frequency. 

The resonator is modeled as a distributed LC circuit consisting of infinitesimal inductances and capacitances extending over the resonator length $l$ [Fig.~\ref{fig:1}(b)].
The circuit's resonant frequency $f_r$ depends on the geometric resonance $f_\mathrm{geo}$ and kinetic contribution $f_\mathrm{kin}$ added in inverse quadrature~\cite{goppl_coplanar_2008},
\begin{equation}
	\label{eq:fr}
	\frac{1}{f_r^2} = \frac{1}{f_\mathrm{geo}^2} + \frac{1}{f_\mathrm{kin}^2},
\end{equation}
where $f_\mathrm{geo}= (2 l \sqrt{ L_\mathrm{geo} C } )^{-1}$ and $f_\mathrm{kin}= (2 l \sqrt{ L_\mathrm{kin} C } )^{-1}$.
The inductance (capacitance) per unit length $L_\mathrm{geo}$ ($C$) is determined by geometry~\cite{supplement}.
In contrast, the kinetic inductance, $L_\mathrm{kin}$, probes the superconducting condensate and has two contributions
\begin{equation}
	\label{eq:fk}
	f_\mathrm{kin}^2 = c_s n_s + c_p n_p,
\end{equation}
where $n_s$ ($n_p$) are normalized superfluid densities associated with the contribution of $s$-wave Al ($p \pm i p$ InAs) superconductors. 
Dimensionless densities  $n_{s,p}$ are normalized to zero-temperature and zero-field limit values, while the parameters $c_s$, $c_p$ encode the zero-temperature and zero-field value of superfluid density and geometry of the sample, thus giving access to the properties of corresponding material. 
The function $n_s$ accounts for the depairing effect of magnetic field and depletion of superfluid density due to thermally activated quasiparticles in Al. 
The function $n_p$ quantifies the depletion of superfluid density in InAs with $p \pm i p$ pairing (assuming high-density and strong spin-orbit coupling, so that interband pairing can be ignored) due to the emergence of Bogoliubov-Fermi surfaces for sufficiently strong magnetic fields. 
We treat the resonator as probing only the $x$-component of $n_p$ because $80\%$ of the resonator is oriented in the $x$-direction.
% $B > \Delta / g \mu_B$, \MS{where $\Delta$ is the gap magnitude and $g$ is the value of $g$-factor.}% or may be remove formula? otherwise we have to explain what's in it
\begin{figure}[t]
	\includegraphics[scale=0.95]{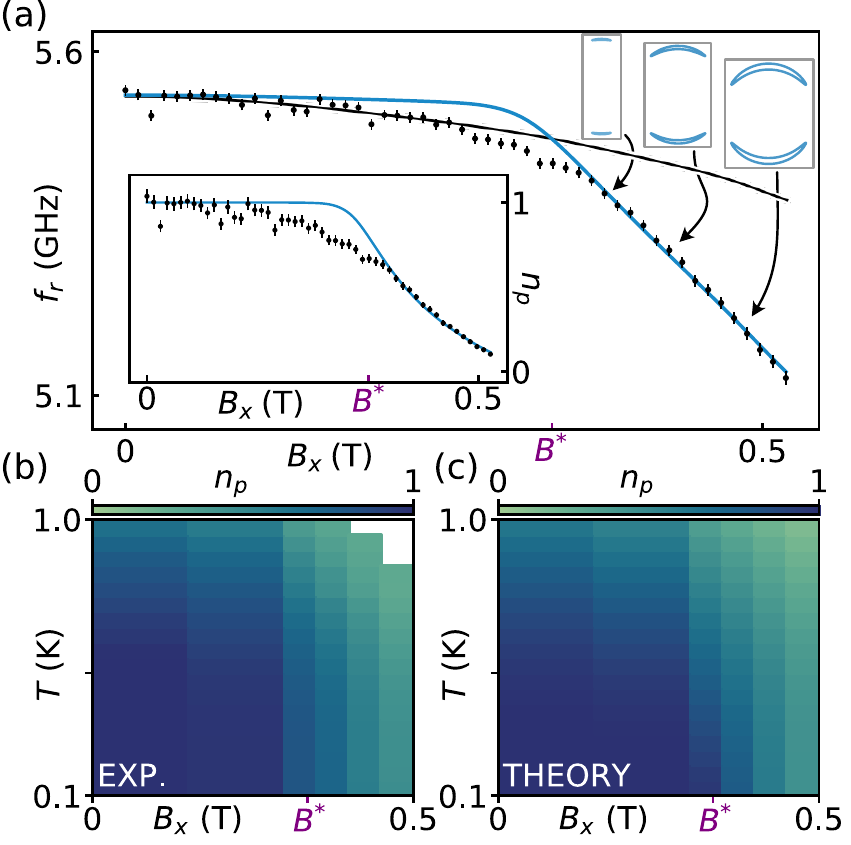}
  	\caption{
	  (a) Points show measured resonant frequency $f_r$ versus $B_x$. 
	  Lines correspond to single-component $s$-wave superconductor model, Eq.~\ref{eq:fk} with $c_p=0$ (black) and to the two-component model with nonzero $c_s$ and $c_p$ (blue).
	Two-component model Bogoliubov-Fermi surfaces are indicated.
	Fit to two-component model is performed by simultaneously fitting linear region of the data for $B>B^*$ and the temperature dependence $f_r(T,B=0)$ in Fig.~\ref{fig:1}(c).
	Inset shows inferred superfluid density $n_p$ with the same $x$-axis as the main figure~\cite{supplement}.
	(b) Experimental inference of $n_p$ versus magnetic field and temperature.
	(c) Theoretical prediction for $n_p$ versus magnetic field and temperature from two-component model.}
	\label{fig:2}
\end{figure}

Microwave access is provided by capacitively coupling the resonator to a transmission line, allowing the transmission coefficient $S_{21}$ to be measured.
The device is placed in a magnetic field with the axis $B_x$ parallel to the transmission line, which can be rotated by an angle $\theta$ in the $B_x$-$B_y$ plane.
A co-fabricated device is used for transport characterization. 
Measuring circuit transmission at $B=0$ and $T=0.1~\mathrm{K}$, a prominent resonance is observed as a dip in the total transmission at a frequency $f_r \approx 5.53\,\mathrm{GHz}$ [Fig.~\ref{fig:1}(b)].
Increasing the cryostat temperature $T$, a frequency down-shift and reduction in quality factor is observed.
As shown in Fig.~\ref{fig:1}(c), the $B=0$ temperature dependence of $f_r$ is nearly identical for the single-component $s$-wave superconductor (solid curve), and a full two-component model that includes the contribution of $p \pm i p$ pairing in the InAs (dashed curve).
To resolve the contribution of the InAs, it is therefore necessary to apply a magnetic field, where one expects large qualitative differences from the standard response of a disordered $s$-wave superconductor (see Fig.~\ref{fig:0}).

The measured dependence of resonant frequency on the value of in-plane field is shown in Fig.~\ref{fig:2}(a), where careful cancellation of perpendicular field was ensured~\cite{supplement}.
Increasing the magnetic field from zero initially causes only a slight decrease in $f_r$, which is qualitatively consistent with the pair-breaking effect of magnetic field on aluminum.
In fact, the black line in Fig.~\ref{fig:2}a shows the prediction of the pair-breaking theory of a single-component $s$-wave superconductor.
This theory utilizes the suppression of critical temperature with in-plane magnetic field, $T_c(B)/T_{c}(0)$, measured on a co-fabricated transport device [Fig.~\ref{fig:1}(e)], and therefore has no free fitting parameters.
Crucially, the resonator response decreases abruptly at a characteristic field scale $B^* \sim 0.33~\mathrm{T}$, in violation of the expectations from pair breaking in pure aluminum.

The decrease of resonator frequency caused by a rapid suppression of superfluid density can be understood by considering the two-component nature of the superconducting condensate.
The model in Eq.~\ref{eq:fk} that incorporates superfluid density contribution of both Al and InAs  is able to adequately capture the full range of frequency behavior (blue line in Fig.~\ref{fig:2}a): it accounts for the conventional behavior at $B < B^{*}$, corresponding to a fully gapped $p$-wave component, and shows the rapid downturn at $B > B^{*}$, corresponding to the emergence of Bogoliubov-Fermi surfaces in the InAs.
The model struggles in the regime $B\sim B^{*}$ because it does not incorporate the role of disorder in the InAs, and therefore underestimates orbital pair-breaking effects.
The rapid onset of the frequency suppression with in-plane field not only provides experimental evidence for the $p\pm ip$ proximity-induced pairing in the InAs semiconductor, but also allows \emph{in situ} characterization of InAs material properties.

Access to material properties is provided via the fit to the theoretical model.
The fit geometric resonant frequency $f_\mathrm{geo}$ is $5.96 \pm 0.01~\mathrm{GHz}$ which differs by $<2~\%$ with  the expected value based on electromagnetic simulations and provides a strong consistency check.
$c_s$ gives the Al sheet resistance $R_\mathrm{Al}=6.7\pm 0.2~\mathrm{\Omega}$, in line with independent transport measurements on MBE-grown Al thin films.
$c_p$ gives the InAs density, $4\times10^{13}~\mathrm{cm^{-2}}$, which combined with the total measured sheet resistance yields an InAs mobility of $2\times10^{4}~\mathrm{cm^2 / (V s)}$.
The density is an order of magnitude larger than without Al, as expected due to band bending of InAs~\cite{reeg_metallization_2018,reeg_proximity_2018,mikkelsen_hybridization_2018,antipov_effects_2018,kiendl_proximity-induced_2019}, whereas the mobility is comparable to the Hall value.
$B^*$ gives a bulk $g$-factor in the $x$-direction, $g_x=11.2 \pm 0.2$, which is consistent with measured $g$-factors in similar quantum wells~\cite{moller_spin_2003,junsaku_gate-controlled_2003,ya_determination_2017,yuan_experimental_2020}.
These parameters give key independent information on the proximity effect in InAs.
% The data are consistent with a large increase in density due to the Al, while maintaining a large induced gap, weak $g$-factor renormalization.
Fermi velocity mismatch between the InAs and Al ($v_{F,\mathrm{InAs}}/v_{F,\mathrm{Al}}\sim 3$) results in a moderate interface transparency with weak $g$-factor renormalization while maintaining a large induced gap due to disorder in the Al~\cite{kiendl_proximity-induced_2019}.
Incorporation of disorder in the InAs will cause quantitative corrections to quantities inferred from the fit.

With all parameters fixed, the normalized $p$-wave superfluid density can now be extracted directly from measured frequencies (Fig.~\ref{fig:2}(a) inset).
Experimentally mapping out a phase diagram for $n_p$ in the $B_x$-$T$ plane reveals that the superfluid density is depleted both by increasing the field above $B^*$, and by raising the temperature, in line with the theoretical model that accounts for both thermal effects and depairing in the $p$-wave system [Fig.~\ref{fig:2}(b),(c)].
This comparison has no free parameters, which provides further strong evidence in favor of the $p \pm i p$ theory.

\begin{figure}[t]
	\centering
	  \includegraphics[scale=1.0]{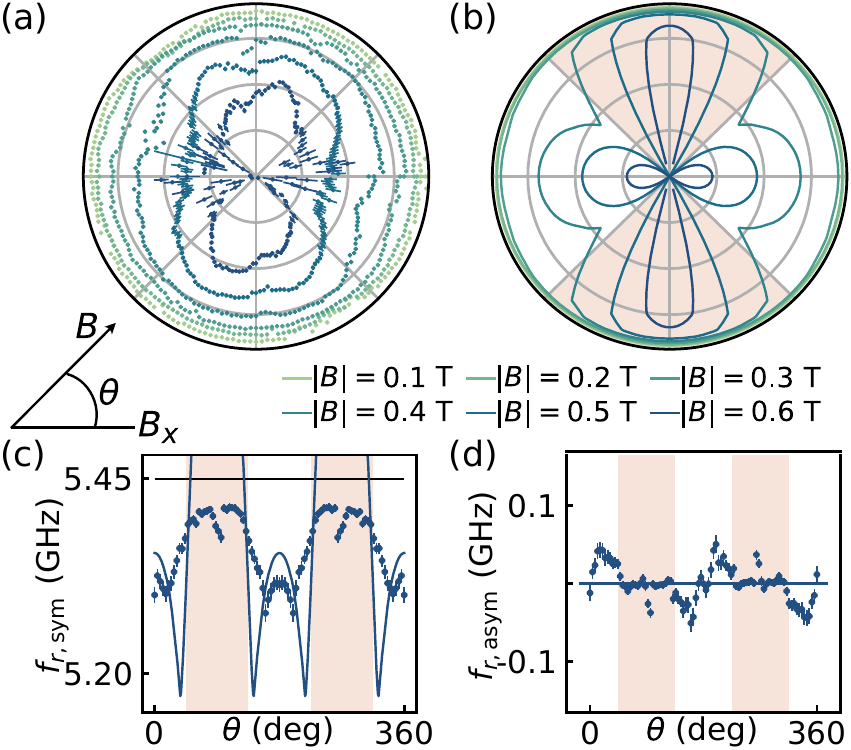}
		\caption{
		(a) Measured magnetic-field orientation dependence of resonant frequency in a polar plot. Radial divisions start at $f_r=4.8~\mathrm{GHz}$ and are in $100~\mathrm{MHz}$ increments.
		(b) Theoretically predicted dependence of frequency on magnetic field orientation, with single free parameter fixed in (c).
		(c) Measured $f_r$ at $|B|=0.4~\mathrm{T}$ symmetrized about $\theta=180^{\circ}$.
		Shaded regions indicate angles for which instability can play a role, which are excluded from the fit for $g_y$.
		Black curve is prediction of the $s$-wave pair-breaking theory.
		Blue curve is a single-parameter fit for $B^*_y$ in the two-component model, which exceeds the value of the pair-breaking model because it does not include disorder.
		(d) Antisymmetric part of $f_r$ vs magnetic field angle $\theta$, blue curve is the theoretical expectation $f_\mathrm{r,asym}=0$.}
	  \label{fig:3}
  \end{figure}

Motivated by the anisotropic nature of the Bogoliubov-Fermi surfaces, we have systematically studied the anisotropy of the circuit response with respect to field direction in Fig.~\ref{fig:3}(a).
Measuring resonant frequency $f_r$ as a function of field angle $\theta$ reveals nearly isotropic response for weak values of magnetic field $B<B^*$.
In contrast, for $B>B^*$ we observe strong frequency suppression in the $x$-direction compared to the $y$-direction, resulting in a pronounced two-lobe structure in a polar frequency plot [Fig.~\ref{fig:3}(a)], with the two prominent lobes at $\theta = \pm 90^\circ$.
There is an additional hint of two smaller lobes at $\theta=0^\circ, 180^\circ$.

In order to compare the measured field-direction-dependence of $f_r$ with theory, we extend our model to include a $g$-factor in the $y$-direction $g_y$ which is expected to %strongly 
differ from $g_x$ for the present case of an asymmetric (100) quantum wells~\cite{kalevich_anisotropy_1992,kalevich_anisotropy_1993,eldridge_spin-orbit_2011}.
Holding all other parameters of the theoretical model fixed, a single-parameter fit in Fig.~\ref{fig:3}(c) yields
a value $g_y = 4$~\cite{supplement}, consistent with the expected level of in-plane $g$-factor anisotropy~\cite{kalevich_anisotropy_1992,kalevich_anisotropy_1993,eldridge_spin-orbit_2011}, and with literature values of in similar quantum wells of $g$-factors in the range of 3-11~\cite{moller_spin_2003,junsaku_gate-controlled_2003,ya_determination_2017,yuan_experimental_2020}.
Remarkably, the addition of this single extra parameter explains the key observed anisotropic features in the dataset [Fig.~\ref{fig:3}(b)].
In particular, theory predicts two major lobes for $B>B^*$, associated with a regime where no Bogoliubov-Fermi arcs emerge due to the relatively small value of $g_y$, and two minor lobes associated with the dependence of arc orientation on field direction.
Both the major and minor lobes predicted by the theoretical model are more prominent than those in the experiment, which we attribute primarily to the theory being in the clean limit, which tends to result in an overestimated frequency for $B<B^*$, as is already apparent in Fig.~\ref{fig:2}(a).
In addition in this sample there is a small odd-angle contribution to the resonant frequency [Fig.~\ref{fig:3}(d)], which is completely absent in theory.
Further experimental work is needed to see if the small odd-angle contribution is experimentally robust.

Interestingly, the semiclassical model predicts that a magnetic field oriented near the $y$-direction (shaded regions in Fig.~\ref{fig:3}), which causes the emergence of Bogoliubov-Fermi arcs aligned with the primary ($x$) direction of the resonator, can result in negative values of the $p\pm ip$ superfluid density ($n_p<0$).
Physically, the negative superfluid density emerges from the large density of quasiparticles due to presence of Bogoliubov-Fermi arcs, and signals an instability which has been discussed in related contexts~\cite{liu_interior_2003,wu_superfluidity_2003,forbes_stability_2005,agterberg_bogoliubov_2017,setty_bogoliubov_2020}. We do not observe indications of the instability in the experiment, possibly due to the presence of Al layer and smearing of density of states in InAs due to disorder. 
The consequences of the instability for the present system remain to be understood and will be addressed in the future work.
In practice, the relatively small value of $g_y$ obtained from the fit leads to the absence of Bogoliubov-Fermi arcs in this region, which effectively masks the role of the unstable region for current experimental parameters.

\begin{figure}
	\centering
	  \includegraphics[scale=0.95]{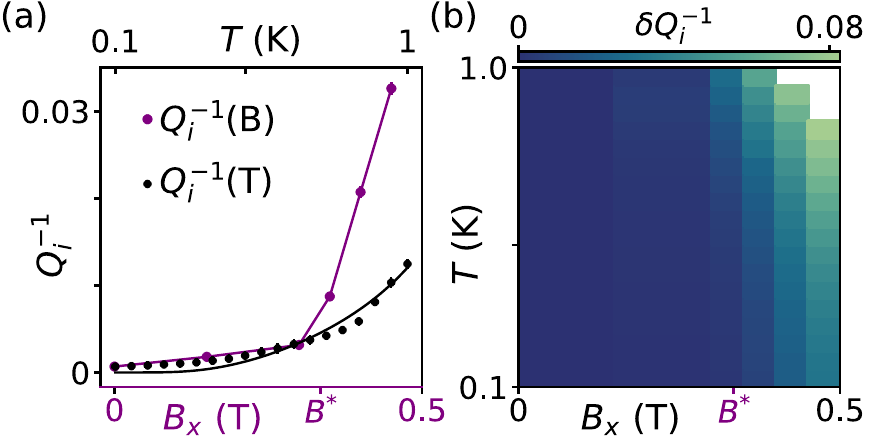}
		\caption{ 
			(a) Dependence of inverse quality factor on magnetic field $Q_i^{-1}(B)$ at a fixed $T=0.1~\mathrm{K}$ (purple), and dependence on temperature $Q_i^{-1}(T)$ at a fixed $B=0$.
	  		Black curve is the zero-field Mattis-Bardeen expectation with no free parameters~\cite{supplement}.
	  		(b) Relative dissipation, $\delta Q_i^{-1}$, as a function of of magnetic field and temperature. $B^*$ indicates the inferred crossover to Bogoliubov-Fermi arcs from Fig.~\ref{fig:2}.}
	  \label{fig:4}
  \end{figure}

To test the origin of the anisotropic response, we have fabricated two additional samples on $90^{\circ}$ rotated crystal axes, and found that the anisotropic circuit response is $90^{\circ}$ rotated as well~\cite{supplement}.
This shows that the origin of anisotropy is associated with the crystal, consistent with an anisotropic $g$-tensor.
%Qualitatively, we speculate that other explanations are also possible, such as the inclusion of Dresselhaus component of the spin-orbit coupling.
It is not currently possible to quantitatively study these orientations because the $90^\circ$ rotated devices strongly sample the unstable region in the currently available $p \pm i p$ theory.
Thus, constructing a more general theory of the induced $p \pm i p$ superfluid response is an outstanding theoretical challenge which must be overcome to analyze all sample orientations.

%%%% FIG 4 discussion %%%%%
An additional check of the $p \pm i p$ picture is given by circuit dissipation.
The circuit's inverse quality-factor $Q_i^{-1}$ increases abruptly at the characteristic field $B^*$ [Fig.~\ref{fig:4}(a)], signaling the onset of enhanced dissipation.
Currently available theory does not include disorder in the InAs, so is unable to make predictions for dissipation signatures of Bogoliubov-Fermi arcs.
We therefore introduce the model-independent dissipation metric
\begin{equation}
	\delta Q_i^{-1}( B, T) = Q_i(B,T)^{-1} - Q_i(0,T)^{-1}.
\end{equation}
$\delta Q_i^{-1}$ represents an inference of the enhanced dissipation due to magnetic field, covering both the high-temperature limit where $Q_i(0,T)^{-1}$ approaches the Mattis-Bardeen prediction [Fig.~\ref{fig:4}(a), black], and the low temperature limit where $Q_i(0,T)^{-1}$ saturates, presumably due to generic effects such as material imperfections.
Experimentally mapping $\delta Q_i^{-1}$ as a function of magnetic field and temperature confirms that there is a generic increase in dissipation for $B>B^*$.
The behavior of $\delta Q_i^{-1}$ bares a striking resemblance to the behavior of $n_p$ in Fig.~\ref{fig:2}(c,d), suggesting the straightforward physical interpretation that the emergence of Bogoliubov-Fermi arcs introduces excess dissipation in the resonator.
Such dissipation is different from the usual Fermi liquid, since carriers have a continuously variable charge which depends on both their momentum and magnetic field, highlighting the need for development of a theoretical description.

Summarizing, we have studied the magnetic field and temperature dependence of an Al-InAs superconducting resonator, observing strong departures from the $s$-wave theory, and good agreement with a theory including the effect of $p \pm i p$ induced superconductivity in the InAs.
Within this picture, a sufficiently strong magnetic field induces anisotropic response and leads to emergence of   Bogoliubov-Fermi surfaces, which result in a rapid shift of the frequency of the resonator and cause sharp onset of excess dissipation.
We have considered other origins of the decreased superfluid density.
A pure induced $s$-wave pairing in the InAs is unable to account for our observations~\cite{supplement}.
Another scenario is that, despite the careful magnetic-field alignment and lack of contribution below $B^*$, there is a depinning transition of vortices \footnote{Al is a type-I superconductor, but in the thin film limit has a field-dependent behavior similar to a type-II~\cite{song_microwave_2009}}.
The absence of extra frequency shifts below $B^{*}$, the weak temperature dependence above $B^{*}$~\cite{prozorov_magnetic_2006}, and the isotropic response of control samples without an InAs heterostructure~\cite{supplement} all point against this scenario.
Anomalous field dependence has been observed in Nb thin films in prior work~\cite{allison_superconducting_2010}; we have verified this is not the case for our films~\cite{supplement}.

Looking ahead, our technique can now be used to study the properties of different hybrid systems, and to explore alternative geometries that could use Bogoliubov-Fermi surfaces to generate topological phases~\cite{papaj_creating_2021}.
% Our evidence of Bogoliubov-Fermi surfaces, and a related result in $\mathrm{Bi_2Te_3}$ that we became aware of during manuscript preparation~\cite{zhen_discovery_2021}, suggest that this is a promising direction for future study.
During preparation we became aware of a related result reporting Bogoliubov-Fermi surfaces~\cite{zhen_discovery_2021}.
%Notably, the emergence of Bogoliubov Fermi surfaces is the first step in a recently-proposed route to a topological phase~\cite{papaj_creating_2021}.
%During the preparation of this manuscript, we became aware of a related work that also reports evidence of Bogoliubov Fermi surfaces in $\mathrm{Bi_2Te_3}$ thin films~\cite{zhen_discovery_2021}.

\begin{acknowledgments}
Maksym Serbyn acknowledges useful discussions with A.~Levchenko, P.~A.~Lee, and E.~Berg.
This research was supported by the Scientific Service Units of IST Austria through resources provided by the MIBA Machine Shop and the nanofabrication facility.
J. Senior and A. Ghazaryan acknowledge funding from the European Union’s Horizon 2020 research and innovation program under the Marie Skłodowska-Curie Grant Agreement No. 754411.W
M. Hatefipour, W.M. Strickland and J. Shabani acknowledge funding from Office of Naval Research award number N00014-21-1-2450.
\end{acknowledgments}

% \bibliography{mesa_res_merged}

	
\newcommand{\suppage}[1]{
	\pagebreak
	\begin{figure}[p]
		\vspace*{-1.5cm}
		\hspace*{-1.9cm}
		\includegraphics[page=#1]{supplement.pdf}
		\centering
	\end{figure}
}

\suppage{1}
\suppage{2}
\suppage{3}
\suppage{4}
\suppage{5}
\suppage{6}
\suppage{7}
\suppage{8}
\suppage{9}
\suppage{10}
\suppage{11}
\suppage{12}
\suppage{13}
\suppage{14}
\suppage{15}
\suppage{16}
\suppage{17}
\suppage{18}
\suppage{19}
\suppage{20}

\end{document}